\documentclass[prl,twocolumn,amsmath,amssymb,superscriptaddress,nofootinbib,10pt]{revtex4-2}
\usepackage[T1]{fontenc}
\usepackage{lmodern}
\usepackage{amsmath}
\usepackage{amsthm}
\usepackage{amsfonts}
\usepackage{xcolor}
\usepackage{slashed}
\usepackage{mathtools}
\usepackage{comment}
\usepackage{epsfig}
\usepackage[colorlinks=true,citecolor=magenta]{hyperref}

\usepackage{url}

\def\ba{\begin{eqnarray}}
\def\ea{\end{eqnarray}}

\def\be#1\ee{\begin{eqnarray}#1\end{eqnarray}}

\def\d{\mathrm{d}}

\def\({\left(}
\def\){\right)}

\def\L{\mathcal{L}}

\newcommand{\nl}{\nonumber \\}

\begin{document}

\title{Graviton propagation in ghost-free massive gravity}

\author{Claudia de Rham}
\email{c.de-rham@imperial.ac.uk}
\affiliation{Abdus Salam Centre for Theoretical Physics, Imperial College, London SW7 2AZ, United Kingdom}

\author{Jan Ko$\dot{\mathrm{z}}$uszek}
\email{jankozuszek@fqa.ub.edu}
\affiliation{Departament de Física Quàntica i Astrofísica and Institut de Ciències del Cosmos (ICC), Universitat de Barcelona, Martí i Franquès 1, ES-08028, Barcelona, Spain.}

\author{Andrew J. Tolley}
\email{a.tolley@imperial.ac.uk}
\affiliation{Abdus Salam Centre for Theoretical Physics, Imperial College, London SW7 2AZ, United Kingdom}

\author{Toby Wiseman}
\email{t.wiseman@imperial.ac.uk}
\affiliation{Abdus Salam Centre for Theoretical Physics, Imperial College, London SW7 2AZ, United Kingdom}


\begin{abstract}

We consider the ghost-free dRGT massive gravity  with two of its three possible mass terms.
This theory has five gravitational degrees of freedom.
On Minkowski spacetime these modes have helicity-2, -1 and -0 and propagate on the Minkowski lightcone in the high-frequency limit.
However for a general background the degrees of freedom corresponding to the helicity-1 and -0 modes have characteristics different to that of the metric lightcone.
Here we prove in all generality, that the two degrees of freedom corresponding to the helicity-2 mode always propagate on the metric lightcone for \emph{any} background in the high-frequency limit, which has significant relevance for current and future observational tests of the theory.

\end{abstract}
\maketitle

\section{Introduction}
\vspace{-0.25cm}

Current observations already place strong constraints on the propagation of gravitational waves. 
For instance, for scalar-tensor theories of gravity, such as Horndeski~\cite{Horndeski:1974wa}, the GW170817 event~\cite{LIGOScientific:2017zic} very tightly constrains either the parameter space~\cite{Creminelli:2017sry,Baker_2017} or the cutoff~\cite{deRham:2018red} of the theory, since generally the helicity-2 mode of the graviton does not propagate on the lightcone of the metric. Those constraints are expected to improve significantly with future multi-messenger and multi-band observations \cite{Baker:2022eiz}. 

Here we consider the dRGT ghost-free massive gravity theory in four dimensions with the standard Einstein-Hilbert kinetic structure, with Minkowski reference metric, and with two of its possible three mass terms. This theory is believed to exhibit a Vainshtein mechanism for an appropriate range of these masses that circumvents the vDVZ discontinuity and potentially allows realistic phenomenology~\cite{Sbisa:2012zk,Koyama:2011yg, Koyama:2011xz,Tasinato:2013rza,Comelli:2011wq,Berezhiani:2013dw, Berezhiani:2011mt,Gruzinov:2011mm,Deffayet:2008zz,Babichev:2009us,Babichev:2010jd,
Chkareuli:2011te,Babichev:2013usa}, although beyond weak field gravity relatively little is known about its behaviour. For example, whether black holes form as the end state of gravitational collapse remains unclear~\cite{Deffayet:2011rh, Berens:2021tzd,Rosen:2017dvn,deRham:2023ngf}.

For the vacuum Minkowski spacetime all helicities of the massive graviton propagate on the lightcone~\cite{deRham:2010kj,deRham:2011qq}. However, this need not be true for general backgrounds. 
In particular, the degrees of freedom corresponding to helicity-0 and -1 modes generally have different characteristics than that of light, as already seen in the simpler decoupling limit \cite{deRham:2010ik,Ondo:2013wka}, although the dynamics of these modes is not  constrained well by current observations~\cite{deRham:2016nuf}.
The question is then; what are the characteristics of the helicity-2 mode for a general background such as might describe our universe?

The decoupling limit of dRGT gravity gives an indication. This is a double scaling limit of small graviton mass and small energy scales compared to the Planck scale $M_{\rm Pl}=1/\sqrt{8\pi G_N}$ (i.e. the weak field limit), and one finds the characteristics for the helicity-2 mode are exactly given by the lightcone of the metric. 
However, the decoupling limit is just the leading term in an expansion in higher derivative interactions, and the well-known subtleties that arise in the massless limit of the theory could imply that this special structure only holds for general backgrounds at leading order in the decoupling limit, even though for special backgrounds such as Minkowski spacetime it holds to all orders.
Indeed, the separation of modes into helicities is only possible on backgrounds with sufficient symmetry, and does not hold in general.

One might expect corrections to the helicity-2 lightcone that depend on the mass and background geometry in such a way that they cancel for Minkowski spacetime and vanish in the decoupling limit. 
This is precisely what happens with Horndeski theories which share the same class of decoupling limit as massive gravity \cite{deRham:2010ik}, where gravitational waves propagate with characteristics that differ from the metric lightcone by operators that may scale as $G_{\mu\nu}/(m M_{\rm Pl} )$ leading to the strong constraints from the GW170817 event.
While corrections of this precise form are not expected for massive gravity, there has previously been no proof that more subtle corrections could not arise that generate sizeable effects, potentially even constraining the Vainshtein mechanism. 

In this Letter we resolve this question. We show the surprising result, of  profound relevance for future observational tests of the theory, that on a fully general background there are two degrees of freedom of the graviton, corresponding to helicity-2 modes on  symmetric backgrounds, that propagate exactly on the metric lightcone at high frequency\footnote{
Recently the same result was shown for the minimal massive gravity~\cite{Kozuszek:2024vyb}. However this minimal theory has a simpler structure and is not capable of a realistic phenomenology, and in particular does not allow for a Vainshtein mechanism~\cite{Koyama:2011yg}.}.

\vspace{-0.25cm}
\section{Ghost-free massive gravity}
\vspace{-0.25cm}

The ghost-free dRGT theory is constructed from a metric $g_{\mu\nu}$ that we assume matter minimally couples to. When we refer to the lightcone, we mean the null cone defined by this metric. The theory also requires  a reference metric $f_{\mu\nu}$ which we will take to be Minkowski spacetime as a natural choice.
We will choose the coordinates so that $f_{\mu\nu} = \eta_{\mu\nu}$. We will also fix the cosmological terms so that the theory admits a Minkowski vacuum.
We construct the theory using a symmetric vielbein $e_{\mu\nu} = e_{(\mu\nu)}$
\cite{Hinterbichler:2012cn}  defined from the metric and reference as
$g_{\mu\nu} = \eta^{\alpha\beta} e_{\alpha\mu}  e_{\beta\nu}$. Then the gravitational part of the action of the theory is
\begin{align}
& S_{g} =  \int \d^4x \sqrt{-g} \left\{\frac{M_{\rm Pl}^2}{2} \left(R 
-\mathcal{U}_{\rm mass}
\right) \nonumber
+\L_{\rm mat}(g_{\mu\nu}, \psi_i)\right\}
 \, \\
& \mathcal{U}_{\rm mass} =  m^2 (2[e]-6) + M^2\(\frac{[e]^2 - [e^2]}{2}-3\)  \,.
\end{align}
Here square brackets represent the trace taken with the dynamical metric, {\it i.e.} $[ e ] = g^{\mu\nu}e_{\mu\nu} = e^\mu_{~\mu}$ 
and $[ e^2 ] = e^\mu_{~\nu}e^\nu_{~\mu}$, and we see that the mass terms are linear and quadratic in the matrix $e^\mu_{~\nu}$. There is a third possible mass term which is cubic which we omit here. As usual, to this action we add the matter action $S^{\rm (matter)}[g]$ where the matter minimally couples to the metric. 

The dRGT theory is characterised by its subtle structure of constraints. We will write the Einstein equation as $\mathcal{R}_{\mu\nu} = 0$ where,
\begin{align}
\mathcal{R}_{\mu\nu} &= R_{\mu\nu} + \mathcal{M}_{\mu\nu} - \kappa^2 \tilde{T}_{\mu\nu} \\
 \mathcal{M}_{\mu\nu} & = m^2 \left( - e_{\mu\nu} - g_{\mu\nu} \left( \frac{[ e ]}{2} - 3 \right) \right) \\
& + \frac{M^2}{2} \left( - [e] e_{\mu\nu} + \eta_{\mu\nu} + 3 g_{\mu\nu} \right) \,, \nonumber
\end{align}
with $\tilde{T}_{\mu\nu}$ the trace reverse of the conserved matter stress tensor and $\kappa=1/M_{\rm Pl}$. Defining $\mathcal{E}_{\mu\nu} =\mathcal{R}_{\mu\nu} - \frac{1}{2} g_{\mu\nu} \mathcal{R}$ then the vector constraint $\xi_\mu = 0$ and the scalar constraint $\mathcal{S} = 0$ are \cite{deRham:2023ngf}
\be
\label{eq:constraints}
  \xi_\mu = - \frac{1}{\mu^2} ( e^{-1} )_{\mu}^{\nu} \nabla^\sigma \mathcal{E}_{\sigma \nu} \; , \quad \mathcal{S} = \frac{1}{2} A^{\mu\nu} \mathcal{R}_{\mu\nu} + \mu^2 \nabla \cdot \xi 
\ee
where
\be
A^{\mu\nu} &=& \left( m^2 + \frac{M^2}{2} [e] \right) g^{\mu\nu} - M^2 e^{\mu\nu}\,.
\ee
Both the vector and scalar constraints involve at most first derivatives of $e_{\mu\nu}$.
For the cubic mass term the scalar constraint cannot be written in an analogous form~\cite{Deser:2014hga}, and together with its poor phenomenology~\cite{Chkareuli:2011te,Berezhiani:2013dca} is the reason we do not consider that case further here. 
On a Minkowski spacetime the graviton 
has mass $\mu^2 = m^2 + M^2$ and propagates on the Minkowski lightcone.
In the following we will choose units so that this flat space graviton mass, $\mu^2 = 1$.

\vspace{-0.25cm}
\section{First order formulation}
\vspace{-0.25cm}

Rather than working with $\mathcal{R}_{\mu\nu}$ we adopt instead a `harmonic' formulation, defined by
\be
\mathcal{R}^H_{\mu\nu} = \mathcal{R}_{\mu\nu} + 2 \nabla_{(\mu} \xi_{\nu)}\,,
\ee
so that the condition $\mathcal{R}^H_{\mu\nu}=0$ promotes the would be degrees of freedom that were previously constrained by $\xi_\mu = 0$ into fully dynamical ones, albeit unphysical constraint-violating modes.
The contracted Bianchi identity then implies
\be
\label{eq:Bianchi}
 \nabla^\mu ( \mathcal{R}^H_{\mu\nu} - \frac{1}{2} g_{\mu\nu} \mathcal{R}^H ) = \nabla^2 \xi_{\nu} + R_{\nu}^{~\mu} \xi_\mu - e_\nu^{~\mu} \xi_\mu
\ee
so that rather than vanishing, now $\xi_\mu$ obeys a wave equation provided the harmonic Einstein equation, $\mathcal{R}^H_{\mu\nu} = 0$ is satisfied.

To proceed we will use this harmonic formulation to write second order equations of motion in a first order form (in all derivatives). To make the argument clearer, we now perform a $3+1$ split of our  coordinates $x^\mu = (t, x^i)=(t,x,y,z)$.

The scalar constraint tells us that one linear combination of the equations $\mathcal{R}^H_{\mu\nu}$ is not second order in derivatives. Hence, instead of solving all the harmonic Einstein equations, we will solve all except $\mathcal{R}^H_{tt}$, and instead we will solve $\partial_t \mathcal{S} = 0$. 
Thus, we must impose the scalar constraint $\mathcal{S} = 0$ on our initial data, which will then ensure that it remains true for later times.
Now, the equations we solve are all of the same second derivative order.

Given that we solve these, the relation~\eqref{eq:Bianchi} implies that the vector constraint $\xi_\mu$ and $\mathcal{R}^H_{tt}$, the equation we do not solve, obey a linear system,
\be
\left(
\left[
\begin{array}{ccc}
\frac{1}{2}A^{tt} \partial_t  &  ( g - A)^{tt} \partial^2_t &  ( g - A)^{ti} \partial^2_t \\
\frac{1}{2}g^{tt} \partial_t  & - g^{tt} \partial^2_t & 0 \\
0 & 0 & -  g^{tt} \partial^2_t
\end{array}
 \right] + \ldots \right)
\left(
\begin{array}{c}
\mathcal{R}^H_{tt} \\
\xi_t \\
\xi_i
\end{array}
\right) = 0 \nl
\ee
where $\ldots$ involve terms with no time derivatives acting on ${R}^H_{tt}$, and at most one when acting on $\xi_\mu$. 
As a consequence,  if initially we have $\mathcal{R}^H_{tt} = 0$ and $\xi_\mu = \partial_t \xi_\nu = 0$, then, provided that $g^{tt} \ne 0$, $\mathcal{R}^H_{tt}  = \xi_\mu = 0$ for all subsequent times.  Setting $\xi_\mu = 0$ imposes the vector constraint on the initial data, just as we had to impose the scalar constraint. The requirement $\dot{\xi}_\mu = 0$ is equivalent to imposing the Hamiltonian and Momentum constraints, associated with the fact the theory is diffeomorphism invariant (although here we choose coordinates so that the reference Minkowski spacetime  takes the canonical form $\eta_{\mu\nu}$).

To proceed to write our second order equations as a first order system in both time and space we must introduce variables for the derivatives of the vierbein. Following our previous work~\cite{deRham:2023ngf,Kozuszek:2024vyb} we take these to be,
\be
K_{\alpha\beta\mu} = \partial_\alpha e_{\beta\mu} -  \partial_\beta e_{\alpha\mu} \; .
\ee
For matter fields, we do the same -- for example, for a canonical scalar field $\Phi$ we would introduce derivative variables $\Psi_\mu = \partial_\mu \psi$.
Then we take the variables for the first order system to be the variables for the matter, together with the 10 independent components of $e_{\mu\nu}$ and 21 of $K_{\alpha\beta\mu}$ which we write as,
\be
\Pi_{i \mu} = K_{t i \mu} \; , \quad Q_{ij\mu} =  K_{ij\mu}
\ee
and noting that $\Pi_{j i} = \Pi_{i j} + Q_{ij t}$ then for $\Pi_{i j}$ we only take independent components with $j \ge i$. Then we may write:
\be
\label{eq:harm}
\mathcal{R}^H_{\mu\nu} &=& \frac{1}{2} \partial_\sigma K_{\alpha\rho\beta} A^{\sigma\alpha\rho\beta}_{\mu\nu} +  B^{\sigma\rho\delta\alpha\beta\gamma}_{\mu\nu} K_{\sigma\rho\delta} K_{\alpha\beta\gamma} \nl
&& \; +  C^{\sigma\rho\delta\alpha\beta\gamma}_{\mu\nu} K_{\sigma\rho\delta} \partial_{(\alpha} e_{\beta)\gamma} + {M}_{\mu\nu} - \kappa \tilde{T}_{\mu\nu}\,.
\ee
The scalar constraint takes the form,
\be
\mathcal{S} = S^{\sigma\rho\delta\alpha\beta\gamma} K_{\sigma\rho\delta} K_{\alpha\beta\gamma} - \tilde{M}\,,
\ee
Its time derivative is then,
\be
\label{eq:scalar}
\partial_t \mathcal{S} &=& 2 ( \partial_t K_{\sigma\rho\delta} ) S^{\sigma\rho\delta\alpha\beta\gamma}  K_{\alpha\beta\gamma} +  \partial_{t} e_{\alpha\beta} H^{\alpha\beta} \nl
&& \qquad +  \partial_{t} e_{\epsilon\eta}   F^{\sigma\rho\delta\alpha\beta\gamma\epsilon\eta} K_{\sigma\rho\delta} K_{\alpha\beta\gamma} \; .
\ee
Crucially, throughout the previous expressions, the tensors $A$, $B$, $C$, $S$, $\mathcal{S}$, $\tilde{M}$  $F$ and $H$  depend only on the veirbein, and not its derivatives. Their explicit forms can be found in the Appendix.

Together the harmonic Einstein equations (except the $tt$ component) and (the time derivative of the) scalar constraint give 10 equations that determine the 10 time derivatives, $\partial_t e_{tt}$ and $\partial_t \Pi_{i \mu}$, and
the system is closed using the following evolution equations for the vierbein and remaining components of the derivative variables, $Q_{ij\mu}$;
\be
\label{eq:evolutioneqns}
\partial_t e_{i \mu}  = K_{t i \mu} + \partial_i e_{t \mu} \; , \quad \partial_t Q_{ij\mu} = \partial_i \Pi_{j \mu} - \partial_j \Pi_{i \mu} \; .
\ee
Altogether this yields  31 equations for the gravity sector that are solved for the time derivatives of the 31 gravitational variables, and in addition the matter sector equations which determine evolution of the matter variables. 
We note that these 1st order variables must obey consistency relations in their initial data, such as {\it e.g.} $\partial_i K_{jk\mu} + \partial_j Q_{ki\mu} + \partial_k Q_{ij\mu} = 0$. Once satisfied on the initial surface, the dynamical equations ensure that these consistency relations hold at all times.

\vspace{-0.25cm}
\section{Linearization and characteristics}
\vspace{-0.25cm}

Now we consider linearizing about a background and we are interested in the characteristics of the fluctuation about that background. Hence we write,
\be
e_{\mu\nu}(x) = \bar{e}_{\mu\nu} +  \epsilon \, e^{\frac{i}{\lambda} k_\alpha(x)  x^\alpha} \delta {e}_{\mu\nu}(x)\,,
\ee
and similarly for the derivative variables $K_{\alpha\beta\mu}$, and matter variables, where we first linearize in $\epsilon$, and then take the high frequency, short wavelength geometric optics limit $\lambda \to 0$, and assume that $k_\alpha(x)$,  $\delta {e}_{\mu\nu}(x)$, $\delta {K}_{\alpha\beta\mu}(x)$ are slowly varying in $x$, and similarly for the matter variables.
Physically this is a wavemode with wavelength much shorter than the physical scales, given by the mass (set to one in out units) and scale of variation of the background metric components.

Then at a point the characteristics are determined by the derivative terms of the fluctuation in the first order system; the non-derivative terms are subdominant in this geometric optics limit. From the harmonic Einstein equations (for $\mu\nu \ne tt$) we find,
\be
\label{eq:linharm}
0 &=& \frac{1}{2} k_\sigma \delta K_{\alpha\rho\beta} \bar{A}^{\sigma\alpha\rho\beta}_{\mu\nu}  +  \bar{C}^{\sigma\rho\delta\alpha\beta\gamma}_{\mu\nu} \bar{K}_{\sigma\rho\delta} k_{(\alpha} \delta e_{\beta)\gamma}  \,,
\ee 
and the time derivative of the scalar constraint implies,
\be
\label{eq:linharm2}
0 &=&  2 \delta K_{\sigma\rho\delta}  \bar{S}^{\sigma\rho\delta\alpha\beta\gamma}  \bar{K}_{\alpha\beta\gamma} +  \delta{e}_{\alpha\beta} \bar{H}^{\alpha\beta}   \nl
&& \qquad +   \delta e_{\epsilon\eta}   \bar{F}^{\sigma\rho\delta\alpha\beta\gamma\epsilon\eta} \bar{K}_{\sigma\rho\delta} \bar{K}_{\alpha\beta\gamma} 
\ee
together with the evolution equations~\eqref{eq:evolutioneqns} which yield,
\be
\label{eq:linevo}
0 & = & k_t \delta e_{i \mu}  - k_i \delta e_{t \mu} \; , \nl
0& = &  k_t \delta Q_{ij\mu} - k_i \delta \Pi_{j \mu} + k_j \delta \Pi_{i \mu}\,.
\ee
The characteristics at a point $x$ are then the solutions of this linear system given by~\eqref{eq:linharm},~\eqref{eq:linharm2} and~\eqref{eq:linevo}, together with the linearized matter equations, for the wavevectors $k_\mu$ together with $\delta {e}_{\mu\nu}$, $\delta \Pi_{i \mu}$ and $\delta Q_{ij\mu}$ and the matter variables at that point. Since we consider matter to be minimally coupled to the metric $g_{\mu\nu}$, the matter variables do not enter the harmonic Einstein equations in the geometric optics limit as  the stress tensor does not have derivatives of the matter variables in -- for example for a massless scalar field ${T}_{\mu\nu} = \Psi_\mu \Psi_\nu - \frac{1}{2} g_{\mu\nu} \Psi^2$.

In order to simplify the analysis we exploit the residual global Lorentz symmetry. 
Consider the point $x$, then we may transform the coordinates so that the background time-space components on the vierbein vanish, $\bar{e}_{ti} = 0$. We may use the residual rotations to align the wavevector at $x$ with the $x$-axis. Thus without loss of generality we may take,
\be
k_\mu = \( -1, \frac{1}{\omega} , 0, 0 \)\,, 
\ee
and consider the characteristics in the $x$-direction. Let us denote the fluctuation components of the vierbein as,
$\delta \phi = \delta e_{tt}$, $\delta V_i = \delta e_{i t}$.
We may reduce the system by noting that the equations~\eqref{eq:linevo} are solved by,
\be
\label{eq:physsol}
\delta V_i = - k_i \delta \phi \, , \; \delta e_{ij} = k_i k_j \delta \phi \, , \; \delta Q_{ij\mu} = 2 k_{[j} \delta \Pi_{i] \mu}\,. \ \ 
\ee
Substituting this into the equations~\eqref{eq:linharm} we obtain a linear system in the 10 variables $\delta \phi$ and $\delta \Pi_{i \mu}$, which is now non-linear in $\omega$ which determines the characteristics of the gravitational degrees of freedom;
\be
\label{eq:gravsystem}
N(\omega) \cdot
\left(
\begin{array}{c}
\delta \phi \\
\delta \Pi_{i t} \\
\delta \Pi_{i j} 
\end{array}
\right) = 0 \; .
\ee
About Minkowski spacetime the determinant of the matrix governing this is $\det(N) \propto (\omega^2 -1)^9$ yielding 9 solutions  with $\omega = \pm 1$, corresponding to 9 wavelike degrees of freedom propagating on the Minkowski lightcone. These comprise the 5 physical massive graviton modes (which can be identified on Minkowski spacetime as 1 scalar, 1 transverse vector, 1 tensor) and 4 constraint violating modes. Let us further divide our coordinates as $x^i = (x, A)$. Then these 9 modes are parameterized by the data $\delta \phi$, $\delta \Pi_{x x}$, $\delta \Pi_{A t}$, $\delta \Pi_{A x}$  and $\delta \Pi_{A B}$ with the remaining $\delta \Pi_{x t}$ determined in terms of these. 

Linearizing the vector constraint $\xi_\mu$ we obtain its fluctuation as,
\be
\delta \xi_\mu = \bar{A}_\mu^{\alpha\beta\sigma} \delta K_{\alpha\beta\sigma} + \bar{K}_{\alpha\beta\sigma} \bar{B}_\mu^{\alpha\beta\sigma\rho\gamma} \delta e_{\rho\gamma}\,,
\ee
with tensors $\bar{A}$, $\bar{B}$ and $\bar{K}$ depending only on the background veirbein.
We note that this only involves the fluctuating variables $\delta K_{\alpha\beta\sigma}$ and $\delta e_{\rho\gamma}$ and not their derivatives.
We may compute the fluctuation in the vector constraint, $\delta \xi_\mu$ about Minkowski spacetime to find,
\begin{align}
&\delta \xi_t = \delta \Pi_{xx} +  \delta \Pi_{AA}  \, , \qquad \delta \xi_x =  \delta \Pi_{xt} - \frac{\delta \Pi_{AA}}{\omega}   \, , \\ 
&  \delta \xi_A =  \delta \Pi_{At} + \frac{\delta \Pi_{A x} }{\omega}\,, \nonumber
\end{align}
where $\omega = \pm 1$. Hence for physical modes with $\delta \xi_\mu = 0$, we may take the 4 variables, $\delta \Pi_{xx}, \delta \Pi_{A t}$ and the trace $\delta \Pi_{A A}$ to be constrained by the remaining data leaving the $9 - 4 = 5$ physical modes.

\vspace{-0.25cm}
\section{Wavemodes on the lightcone}
\vspace{-0.25cm}

Deducing the characteristics of this system is very complicated for a general background. However we will now show relatively straightforwardly that the helicity-2 graviton propagates on the lightcone. After some computation one finds the elegant expressions;
\be
&& \delta \mathcal{R}^H_{AB}  =   \frac{ \bar{k}^2}{2}  \Big[  \left( \delta^i_A (\bar{e}^{-1})^j_{~B} + \delta^i_B (\bar{e}^{-1})^j_{~A} \right) \delta \Pi_{ij} \nl
 && \qquad \qquad + 2\bar{K}_{\sigma \rho \delta}  \left( (\bar{e}^{-1})^{\rho t} - \frac{(\bar{e}^{-1})^{\rho x}}{\omega}   \right)   (\bar{e}^{-1})_{(A}^{\delta} \delta^\sigma_{B)} 
 \delta \phi 
 \Big] \; , \nonumber
 \ee
  \be
&&\delta \mathcal{R}^H_{A t} + \omega  \delta \mathcal{R}^H_{A x}  =  
   \omega \bar{k}^2 \Big[
   (\bar{e}^{-1})^i_{~(A} \delta \Pi_{x) i} + \frac{ (\bar{e}^{-1})^t_{~t} }{2 \omega} \delta \Pi_{A t}  \nl
   &&   \; 
    + 
  \bar{K}_{\sigma \rho \delta} \left( (\bar{e}^{-1})^{\rho t} - \frac{(\bar{e}^{-1})^{\rho x}}{\omega}   \right) (\bar{e}^{-1})_{(A}^{\delta} \Big(   \delta^\sigma_{x)}  
   + \frac{1}{\omega}   \delta^\sigma_{t)}  \Big)
 \delta \phi 
 \Big] \nl
 \ee   
 where we have defined $\bar{k}^2 = k_\mu k_\nu \bar{g}^{\mu\nu}$.
Thus for modes on the lightcone, so that $0= \bar{k}^2 = \frac{1}{\omega^2} \left( \omega^2 \bar{g}^{tt} + \bar{g}^{xx} \right) $ then these 5 equations trivially vanish. Denoting these values of $\omega$ as $\omega^*$ then 
we see that the matrix ${N}(\omega^*)$  governing the gravitational wavemode in~\eqref{eq:gravsystem} has at least a 5 dimensional kernel. Hence there are at least 5 {\it independent} fluctuation wavemodes that propagate on the metric lightcone. However not all of these are physical fluctuations of the original theory. \emph{If} all the components of $\delta \xi_\mu$ were linearly independent for these wavemodes, then imposing the fluctuation is physical would impose 4 conditions on these 5 modes, leaving only one physical mode. However, as we now show on the lightcone not all components of $\delta \xi_\mu$ are independent,  resulting in two of the five wavemodes being physical, and corresponding to the helicity-2 graviton.

\vspace{-0.25cm}
\section{Physical and unphysical wavemodes}
\vspace{-0.25cm}

To understand which of the 5 wavemodes are physical we return to the Bianchi relation~\eqref{eq:Bianchi}, noting that neither $\delta R^H_{tt}$ or the fluctuation in the vector constraint $\delta \xi_\mu$ needs to vanish. Linearizing this about a background satisfying the vector constraint, so $\bar{\xi}_\mu = 0$, gives,
\be
0 =  \delta_{t\nu} \bar{g}^{tt} \partial_t \delta \mathcal{R}^H_{tt} - \frac{1}{2} \bar{g}^{tt} \partial_t \delta \mathcal{R}^H_{tt}  - \partial^\mu \partial_\mu \delta\xi_{\nu}  + \ldots 
\ee
where the terms $\ldots$ are lower order terms in derivatives of the fluctuation; they have no derivatives of $\delta \mathcal{R}^H_{tt}$, only one derivative of $\delta \xi_\mu$, the fluctuation of a Christoffel symbol with no derivatives on it, or one derivative of the veirbein.  Then using the physical solution for derivatives of the vierbein~\eqref{eq:physsol} and focussing again on the characteristics in the $x$ direction (and hence $\partial_t \to -1$ and $\partial_x \to \frac{1}{\omega}$ on the fluctuation), then the four components of the above yield,
\be
\label{eq:eq1}
 - \frac{1}{2} \bar{g}^{tt} \delta \mathcal{R}^H_{tt} =  \bar{k}^2 \delta \xi_t = \omega \bar{k}^2 \delta \xi_x \; , \quad
 0  = \bar{k}^2 \delta \xi_A  \,.
\ee
A similar analysis for the time derivative of the scalar constraint yields
\be
\label{eq:eq2}
 \frac{1}{2} \bar{A}^{tt} \delta \mathcal{R}^H_{tt} =   ( \bar{g}^{tt} - \bar{A}^{tt} )  \delta \xi_t  - \frac{1}{\omega} ( \bar{g}^{xi} - \bar{A}^{xi} ) \delta \xi_i  \; .
\ee
Now, if we consider a mode on the metric lightcone, so that $\bar{k}^2 = 0$, then~\eqref{eq:eq1} implies that $\delta R^H_{tt} = 0$ for that wavemode, and then~\eqref{eq:eq2} implies a linear relation between the components of $\delta \xi_\mu$. Thus, on the lightcone, imposing the physical constraint $\delta \xi_\mu = 0$ only imposes 3 conditions on our wavemode.

Thus in summary, for these 5 modes that travel on the metric lightcone, only three are modes where the constraint fluctuates to be non-zero. The remaining two are physical modes. From the above, it is straightforward to see that a remaining constraint violating fluctuation generically propagates off the lightcone.

\vspace{-0.25cm}
\section{Relation to helicity-2 graviton}
\vspace{-0.25cm}

For a wavemode propagating in the $x$ direction, a general background breaks the rotational symmetry about the $x$-axis. Hence the rotational symmetry underlying helicity is explicitly broken.\footnote{Indeed wavemodes exhibit birefringent propagation in the minimal theory in~\cite{Kozuszek:2024vyb}.} However, we may identify our two lightcone modes by seeing what their behaviour is on a background that does preserve rotations about $x$, but breaks the degeneracy between the different helicities. An example is a background where at a point $p$ we have $\bar{e}_{\mu\nu}|_p = \left(
\begin{array}{cc}
-1 & 0 \\
0 & A \delta_{ij} 
\end{array}
\right)$
and $\bar{K}_{\alpha\beta\mu}|_p =0$, preserving rotational symmetry about the $x$-axis there. 
The vector constraint of the background at the point is satisfied, $\bar{\xi}_\mu|_p = 0$, but the scalar constraint is not. Hence this does not correspond to a physical background for massive gravity, but does allow us to see how the helicity degeneracy splits.
The lightcone at $p$ is determined by $\omega^2 = 1/A^2$.
Then writing $m^2 = 1 - \alpha$, $M^2 = \alpha$, the 5 wavemode fluctuations which preserve the vector constraint, $\delta \xi_\mu = 0$, have;
\begin{center}
\begin{tabular}{c|c|c}
helicity & $\omega^2$  \\
\hline
2 & $ \frac{1}{A^2}$     \\
1 & 
$\frac{ \alpha + 2 A + A^2 (2 - \alpha ) }{4 A^2 \left( A (1 -\alpha ) + \alpha \right)}$
   \\
0 & 
$ \frac{2 A^2 + 4(1-A) A \alpha + (1-A)^2 \alpha^2 }{A^2 ( 2 A^2 + 6(1-A)A \alpha + 3 (1 - A)^2 \alpha^2) }$
 
\end{tabular}
\end{center}
We see the helicity-2 modes propagate on the metric lightcone, corresponding to fluctuations of $\delta \Pi_{yz}$, $\delta \Pi_{yy} - \delta \Pi_{zz}$.
The helicity-0 and -1 modes correspond to fluctuations of $\delta \phi$, and to linear combinations of $\delta \Pi_{A t}$ and $\delta \Pi_{A x}$, and propagate off the metric lightcone for $A \ne 1$.

\vspace{-0.25cm}
\section{Summary}
\vspace{-0.25cm}

In this Letter we have shown that two degrees of freedom of the graviton have characteristics determined by the metric lightcone for dRGT massive gravity with its first two mass terms  on a fully general background. These transform as helicity-2 modes on symmetric backgrounds. 

This is important for observational consistency of the theory and confirms the expectation that there can be no large corrections away from the metric lightcone for the helicity-2 mode.
The decoupling limit of the theory suggested this result, but for Horndeski, which shares the same class of decoupling limit, there are large corrections away from the decoupling limit leading to powerful constraints from the GW170817 observation. 
Not only does our result show that in dRGT there are no such large corrections to the behaviour in the decoupling limit, it shows there are no corrections at all.

A complete understanding of the characteristics of the helicity-1 and -0 modes remains a challenge for this theory due to the existence of highly-non-trivial higher derivative interactions.
Given this, it is remarkable that the helicity-2 graviton, which couples in a highly non-linear manner, has such simple characteristics. This indicates a deeper structure exists within the theory which would be absent if the decoupling limit was differently `covariantised'  or if the helicity-2 and -0 modes were treated as tensor and scalar degrees of freedom in their own right as is for instance the case in Horndeski and its generalisations.

This result is suggestive that one might interpret massive gravity as GR coupled to some effective matter sector, where the “matter” is built out of the helicity‑0 and helicity‑1 modes. This perspective may have implications for possible UV completions of the theory.

\vspace{-0.25cm}
\section*{Acknowledgements}
\vspace{-0.25cm}

The work of CdR, AJT and TW is supported by STFC Consolidated Grant ST/X000575/1. The work of CdR is also supported by Simons Investigator award 690508. JK acknowledges financial support from Grant CEX2024-001451-M funded by MICIU-AEI-10.13039/501100011033, from Grant No. PID2022-136224NB-C22 from the Spanish Ministry of Science, Innovation and Universities, and from Grant No. 2021-SGR-872 funded by the Catalan Government. This research is also funded by the European Union (ERC, HoloGW, Grant Agreement No. 101141909). Views and opinions expressed are, however, those of the authors only and do not necessarily reflect those of the European Union or the European Research Council. Neither the European Union nor the granting authority can be held responsible for them.

\addcontentsline{toc}{section}{Bibliography}
\bibliography{refs}

\begin{thebibliography}{32}%
\makeatletter
\providecommand \@ifxundefined [1]{%
 \@ifx{#1\undefined}
}%
\providecommand \@ifnum [1]{%
 \ifnum #1\expandafter \@firstoftwo
 \else \expandafter \@secondoftwo
 \fi
}%
\providecommand \@ifx [1]{%
 \ifx #1\expandafter \@firstoftwo
 \else \expandafter \@secondoftwo
 \fi
}%
\providecommand \natexlab [1]{#1}%
\providecommand \enquote  [1]{``#1''}%
\providecommand \bibnamefont  [1]{#1}%
\providecommand \bibfnamefont [1]{#1}%
\providecommand \citenamefont [1]{#1}%
\providecommand \href@noop [0]{\@secondoftwo}%
\providecommand \href [0]{\begingroup \@sanitize@url \@href}%
\providecommand \@href[1]{\@@startlink{#1}\@@href}%
\providecommand \@@href[1]{\endgroup#1\@@endlink}%
\providecommand \@sanitize@url [0]{\catcode `\\12\catcode `\$12\catcode
  `\&12\catcode `\#12\catcode `\^12\catcode `\_12\catcode `\%12\relax}%
\providecommand \@@startlink[1]{}%
\providecommand \@@endlink[0]{}%
\providecommand \url  [0]{\begingroup\@sanitize@url \@url }%
\providecommand \@url [1]{\endgroup\@href {#1}{\urlprefix }}%
\providecommand \urlprefix  [0]{URL }%
\providecommand \Eprint [0]{\href }%
\providecommand \doibase [0]{https://doi.org/}%
\providecommand \selectlanguage [0]{\@gobble}%
\providecommand \bibinfo  [0]{\@secondoftwo}%
\providecommand \bibfield  [0]{\@secondoftwo}%
\providecommand \translation [1]{[#1]}%
\providecommand \BibitemOpen [0]{}%
\providecommand \bibitemStop [0]{}%
\providecommand \bibitemNoStop [0]{.\EOS\space}%
\providecommand \EOS [0]{\spacefactor3000\relax}%
\providecommand \BibitemShut  [1]{\csname bibitem#1\endcsname}%
\let\auto@bib@innerbib\@empty
\bibitem [{\citenamefont {Horndeski}(1974)}]{Horndeski:1974wa}%
  \BibitemOpen
  \bibfield  {author} {\bibinfo {author} {\bibfnamefont {G.~W.}\ \bibnamefont
  {Horndeski}},\ }\bibfield  {title} {\bibinfo {title} {{Second-order
  scalar-tensor field equations in a four-dimensional space}},\ }\href
  {https://doi.org/10.1007/BF01807638} {\bibfield  {journal} {\bibinfo
  {journal} {Int. J. Theor. Phys.}\ }\textbf {\bibinfo {volume} {10}},\
  \bibinfo {pages} {363} (\bibinfo {year} {1974})}\BibitemShut {NoStop}%
\bibitem [{\citenamefont {Abbott}\ \emph {et~al.}(2017)\citenamefont {Abbott}
  \emph {et~al.}}]{LIGOScientific:2017zic}%
  \BibitemOpen
  \bibfield  {author} {\bibinfo {author} {\bibfnamefont {B.~P.}\ \bibnamefont
  {Abbott}} \emph {et~al.} (\bibinfo {collaboration} {LIGO Scientific, Virgo,
  Fermi-GBM, INTEGRAL}),\ }\bibfield  {title} {\bibinfo {title} {{Gravitational
  Waves and Gamma-rays from a Binary Neutron Star Merger: GW170817 and GRB
  170817A}},\ }\href {https://doi.org/10.3847/2041-8213/aa920c} {\bibfield
  {journal} {\bibinfo  {journal} {Astrophys. J. Lett.}\ }\textbf {\bibinfo
  {volume} {848}},\ \bibinfo {pages} {L13} (\bibinfo {year} {2017})},\ \Eprint
  {https://arxiv.org/abs/1710.05834} {arXiv:1710.05834 [astro-ph.HE]}
  \BibitemShut {NoStop}%
\bibitem [{\citenamefont {Creminelli}\ and\ \citenamefont
  {Vernizzi}(2017)}]{Creminelli:2017sry}%
  \BibitemOpen
  \bibfield  {author} {\bibinfo {author} {\bibfnamefont {P.}~\bibnamefont
  {Creminelli}}\ and\ \bibinfo {author} {\bibfnamefont {F.}~\bibnamefont
  {Vernizzi}},\ }\bibfield  {title} {\bibinfo {title} {{Dark Energy after
  GW170817 and GRB170817A}},\ }\href
  {https://doi.org/10.1103/PhysRevLett.119.251302} {\bibfield  {journal}
  {\bibinfo  {journal} {Phys. Rev. Lett.}\ }\textbf {\bibinfo {volume} {119}},\
  \bibinfo {pages} {251302} (\bibinfo {year} {2017})},\ \Eprint
  {https://arxiv.org/abs/1710.05877} {arXiv:1710.05877 [astro-ph.CO]}
  \BibitemShut {NoStop}%
\bibitem [{\citenamefont {Baker}\ \emph {et~al.}(2017)\citenamefont {Baker},
  \citenamefont {Bellini}, \citenamefont {Ferreira}, \citenamefont {Lagos},
  \citenamefont {Noller},\ and\ \citenamefont {Sawicki}}]{Baker_2017}%
  \BibitemOpen
  \bibfield  {author} {\bibinfo {author} {\bibfnamefont {T.}~\bibnamefont
  {Baker}}, \bibinfo {author} {\bibfnamefont {E.}~\bibnamefont {Bellini}},
  \bibinfo {author} {\bibfnamefont {P.}~\bibnamefont {Ferreira}}, \bibinfo
  {author} {\bibfnamefont {M.}~\bibnamefont {Lagos}}, \bibinfo {author}
  {\bibfnamefont {J.}~\bibnamefont {Noller}},\ and\ \bibinfo {author}
  {\bibfnamefont {I.}~\bibnamefont {Sawicki}},\ }\bibfield  {title} {\bibinfo
  {title} {Strong constraints on cosmological gravity from gw170817 and grb
  170817a},\ }\bibfield  {journal} {\bibinfo  {journal} {Physical Review
  Letters}\ }\textbf {\bibinfo {volume} {119}},\ \href
  {https://doi.org/10.1103/physrevlett.119.251301}
  {10.1103/physrevlett.119.251301} (\bibinfo {year} {2017})\BibitemShut
  {NoStop}%
\bibitem [{\citenamefont {de~Rham}\ and\ \citenamefont
  {Melville}(2018)}]{deRham:2018red}%
  \BibitemOpen
  \bibfield  {author} {\bibinfo {author} {\bibfnamefont {C.}~\bibnamefont
  {de~Rham}}\ and\ \bibinfo {author} {\bibfnamefont {S.}~\bibnamefont
  {Melville}},\ }\bibfield  {title} {\bibinfo {title} {{Gravitational Rainbows:
  LIGO and Dark Energy at its Cutoff}},\ }\href
  {https://doi.org/10.1103/PhysRevLett.121.221101} {\bibfield  {journal}
  {\bibinfo  {journal} {Phys. Rev. Lett.}\ }\textbf {\bibinfo {volume} {121}},\
  \bibinfo {pages} {221101} (\bibinfo {year} {2018})},\ \Eprint
  {https://arxiv.org/abs/1806.09417} {arXiv:1806.09417 [hep-th]} \BibitemShut
  {NoStop}%
\bibitem [{\citenamefont {Baker}\ \emph {et~al.}(2023)\citenamefont {Baker},
  \citenamefont {Barausse}, \citenamefont {Chen}, \citenamefont {de~Rham},
  \citenamefont {Pieroni},\ and\ \citenamefont {Tasinato}}]{Baker:2022eiz}%
  \BibitemOpen
  \bibfield  {author} {\bibinfo {author} {\bibfnamefont {T.}~\bibnamefont
  {Baker}}, \bibinfo {author} {\bibfnamefont {E.}~\bibnamefont {Barausse}},
  \bibinfo {author} {\bibfnamefont {A.}~\bibnamefont {Chen}}, \bibinfo {author}
  {\bibfnamefont {C.}~\bibnamefont {de~Rham}}, \bibinfo {author} {\bibfnamefont
  {M.}~\bibnamefont {Pieroni}},\ and\ \bibinfo {author} {\bibfnamefont
  {G.}~\bibnamefont {Tasinato}},\ }\bibfield  {title} {\bibinfo {title}
  {{Testing gravitational wave propagation with multiband detections}},\ }\href
  {https://doi.org/10.1088/1475-7516/2023/03/044} {\bibfield  {journal}
  {\bibinfo  {journal} {JCAP}\ }\textbf {\bibinfo {volume} {03}},\ \bibinfo
  {pages} {044}},\ \Eprint {https://arxiv.org/abs/2209.14398} {arXiv:2209.14398
  [gr-qc]} \BibitemShut {NoStop}%
\bibitem [{\citenamefont {Sbisa}\ \emph {et~al.}(2012)\citenamefont {Sbisa},
  \citenamefont {Niz}, \citenamefont {Koyama},\ and\ \citenamefont
  {Tasinato}}]{Sbisa:2012zk}%
  \BibitemOpen
  \bibfield  {author} {\bibinfo {author} {\bibfnamefont {F.}~\bibnamefont
  {Sbisa}}, \bibinfo {author} {\bibfnamefont {G.}~\bibnamefont {Niz}}, \bibinfo
  {author} {\bibfnamefont {K.}~\bibnamefont {Koyama}},\ and\ \bibinfo {author}
  {\bibfnamefont {G.}~\bibnamefont {Tasinato}},\ }\bibfield  {title} {\bibinfo
  {title} {{Characterising Vainshtein Solutions in Massive Gravity}},\ }\href
  {https://doi.org/10.1103/PhysRevD.86.024033} {\bibfield  {journal} {\bibinfo
  {journal} {Phys.Rev.}\ }\textbf {\bibinfo {volume} {D86}},\ \bibinfo {pages}
  {024033} (\bibinfo {year} {2012})},\ \Eprint
  {https://arxiv.org/abs/1204.1193} {arXiv:1204.1193 [hep-th]} \BibitemShut
  {NoStop}%
\bibitem [{\citenamefont {Koyama}\ \emph
  {et~al.}(2011{\natexlab{a}})\citenamefont {Koyama}, \citenamefont {Niz},\
  and\ \citenamefont {Tasinato}}]{Koyama:2011yg}%
  \BibitemOpen
  \bibfield  {author} {\bibinfo {author} {\bibfnamefont {K.}~\bibnamefont
  {Koyama}}, \bibinfo {author} {\bibfnamefont {G.}~\bibnamefont {Niz}},\ and\
  \bibinfo {author} {\bibfnamefont {G.}~\bibnamefont {Tasinato}},\ }\bibfield
  {title} {\bibinfo {title} {{Strong interactions and exact solutions in
  non-linear massive gravity}},\ }\href
  {https://doi.org/10.1103/PhysRevD.84.064033} {\bibfield  {journal} {\bibinfo
  {journal} {Phys.Rev.}\ }\textbf {\bibinfo {volume} {D84}},\ \bibinfo {pages}
  {064033} (\bibinfo {year} {2011}{\natexlab{a}})},\ \Eprint
  {https://arxiv.org/abs/1104.2143} {arXiv:1104.2143 [hep-th]} \BibitemShut
  {NoStop}%
\bibitem [{\citenamefont {Koyama}\ \emph
  {et~al.}(2011{\natexlab{b}})\citenamefont {Koyama}, \citenamefont {Niz},\
  and\ \citenamefont {Tasinato}}]{Koyama:2011xz}%
  \BibitemOpen
  \bibfield  {author} {\bibinfo {author} {\bibfnamefont {K.}~\bibnamefont
  {Koyama}}, \bibinfo {author} {\bibfnamefont {G.}~\bibnamefont {Niz}},\ and\
  \bibinfo {author} {\bibfnamefont {G.}~\bibnamefont {Tasinato}},\ }\bibfield
  {title} {\bibinfo {title} {{Analytic solutions in non-linear massive
  gravity}},\ }\href {https://doi.org/10.1103/PhysRevLett.107.131101}
  {\bibfield  {journal} {\bibinfo  {journal} {Phys.Rev.Lett.}\ }\textbf
  {\bibinfo {volume} {107}},\ \bibinfo {pages} {131101} (\bibinfo {year}
  {2011}{\natexlab{b}})},\ \Eprint {https://arxiv.org/abs/1103.4708}
  {arXiv:1103.4708 [hep-th]} \BibitemShut {NoStop}%
\bibitem [{\citenamefont {Tasinato}\ \emph {et~al.}(2013)\citenamefont
  {Tasinato}, \citenamefont {Koyama},\ and\ \citenamefont
  {Niz}}]{Tasinato:2013rza}%
  \BibitemOpen
  \bibfield  {author} {\bibinfo {author} {\bibfnamefont {G.}~\bibnamefont
  {Tasinato}}, \bibinfo {author} {\bibfnamefont {K.}~\bibnamefont {Koyama}},\
  and\ \bibinfo {author} {\bibfnamefont {G.}~\bibnamefont {Niz}},\ }\bibfield
  {title} {\bibinfo {title} {{Exact Solutions in Massive Gravity}},\ }\href
  {https://doi.org/10.1088/0264-9381/30/18/184002} {\bibfield  {journal}
  {\bibinfo  {journal} {Class.Quant.Grav.}\ }\textbf {\bibinfo {volume} {30}},\
  \bibinfo {pages} {184002} (\bibinfo {year} {2013})},\ \Eprint
  {https://arxiv.org/abs/1304.0601} {arXiv:1304.0601 [hep-th]} \BibitemShut
  {NoStop}%
\bibitem [{\citenamefont {Comelli}\ \emph {et~al.}(2012)\citenamefont
  {Comelli}, \citenamefont {Crisostomi}, \citenamefont {Nesti},\ and\
  \citenamefont {Pilo}}]{Comelli:2011wq}%
  \BibitemOpen
  \bibfield  {author} {\bibinfo {author} {\bibfnamefont {D.}~\bibnamefont
  {Comelli}}, \bibinfo {author} {\bibfnamefont {M.}~\bibnamefont {Crisostomi}},
  \bibinfo {author} {\bibfnamefont {F.}~\bibnamefont {Nesti}},\ and\ \bibinfo
  {author} {\bibfnamefont {L.}~\bibnamefont {Pilo}},\ }\bibfield  {title}
  {\bibinfo {title} {{Spherically Symmetric Solutions in Ghost-Free Massive
  Gravity}},\ }\href {https://doi.org/10.1103/PhysRevD.85.024044} {\bibfield
  {journal} {\bibinfo  {journal} {Phys.Rev.}\ }\textbf {\bibinfo {volume}
  {D85}},\ \bibinfo {pages} {024044} (\bibinfo {year} {2012})},\ \Eprint
  {https://arxiv.org/abs/1110.4967} {arXiv:1110.4967 [hep-th]} \BibitemShut
  {NoStop}%
\bibitem [{\citenamefont {Berezhiani}\ \emph
  {et~al.}(2013{\natexlab{a}})\citenamefont {Berezhiani}, \citenamefont
  {Chkareuli},\ and\ \citenamefont {Gabadadze}}]{Berezhiani:2013dw}%
  \BibitemOpen
  \bibfield  {author} {\bibinfo {author} {\bibfnamefont {L.}~\bibnamefont
  {Berezhiani}}, \bibinfo {author} {\bibfnamefont {G.}~\bibnamefont
  {Chkareuli}},\ and\ \bibinfo {author} {\bibfnamefont {G.}~\bibnamefont
  {Gabadadze}},\ }\bibfield  {title} {\bibinfo {title} {{Restricted
  Galileons}},\ }\href {https://doi.org/10.1103/PhysRevD.88.124020} {\bibfield
  {journal} {\bibinfo  {journal} {Phys. Rev. D}\ }\textbf {\bibinfo {volume}
  {88}},\ \bibinfo {pages} {124020} (\bibinfo {year} {2013}{\natexlab{a}})},\
  \Eprint {https://arxiv.org/abs/1302.0549} {arXiv:1302.0549 [hep-th]}
  \BibitemShut {NoStop}%
\bibitem [{\citenamefont {Berezhiani}\ \emph {et~al.}(2012)\citenamefont
  {Berezhiani}, \citenamefont {Chkareuli}, \citenamefont {de~Rham},
  \citenamefont {Gabadadze},\ and\ \citenamefont {Tolley}}]{Berezhiani:2011mt}%
  \BibitemOpen
  \bibfield  {author} {\bibinfo {author} {\bibfnamefont {L.}~\bibnamefont
  {Berezhiani}}, \bibinfo {author} {\bibfnamefont {G.}~\bibnamefont
  {Chkareuli}}, \bibinfo {author} {\bibfnamefont {C.}~\bibnamefont {de~Rham}},
  \bibinfo {author} {\bibfnamefont {G.}~\bibnamefont {Gabadadze}},\ and\
  \bibinfo {author} {\bibfnamefont {A.}~\bibnamefont {Tolley}},\ }\bibfield
  {title} {\bibinfo {title} {{On Black Holes in Massive Gravity}},\ }\href
  {https://doi.org/10.1103/PhysRevD.85.044024} {\bibfield  {journal} {\bibinfo
  {journal} {Phys.Rev.}\ }\textbf {\bibinfo {volume} {D85}},\ \bibinfo {pages}
  {044024} (\bibinfo {year} {2012})},\ \Eprint
  {https://arxiv.org/abs/1111.3613} {arXiv:1111.3613 [hep-th]} \BibitemShut
  {NoStop}%
\bibitem [{\citenamefont {Gruzinov}\ and\ \citenamefont
  {Mirbabayi}(2011)}]{Gruzinov:2011mm}%
  \BibitemOpen
  \bibfield  {author} {\bibinfo {author} {\bibfnamefont {A.}~\bibnamefont
  {Gruzinov}}\ and\ \bibinfo {author} {\bibfnamefont {M.}~\bibnamefont
  {Mirbabayi}},\ }\bibfield  {title} {\bibinfo {title} {{Stars and Black Holes
  in Massive Gravity}},\ }\href {https://doi.org/10.1103/PhysRevD.84.124019}
  {\bibfield  {journal} {\bibinfo  {journal} {Phys.Rev.}\ }\textbf {\bibinfo
  {volume} {D84}},\ \bibinfo {pages} {124019} (\bibinfo {year} {2011})},\
  \Eprint {https://arxiv.org/abs/1106.2551} {arXiv:1106.2551 [hep-th]}
  \BibitemShut {NoStop}%
\bibitem [{\citenamefont {Deffayet}(2008)}]{Deffayet:2008zz}%
  \BibitemOpen
  \bibfield  {author} {\bibinfo {author} {\bibfnamefont {C.}~\bibnamefont
  {Deffayet}},\ }\bibfield  {title} {\bibinfo {title} {{Spherically symmetric
  solutions of massive gravity}},\ }\href
  {https://doi.org/10.1088/0264-9381/25/15/154007} {\bibfield  {journal}
  {\bibinfo  {journal} {Class.Quant.Grav.}\ }\textbf {\bibinfo {volume} {25}},\
  \bibinfo {pages} {154007} (\bibinfo {year} {2008})}\BibitemShut {NoStop}%
\bibitem [{\citenamefont {Babichev}\ \emph {et~al.}(2009)\citenamefont
  {Babichev}, \citenamefont {Deffayet},\ and\ \citenamefont
  {Ziour}}]{Babichev:2009us}%
  \BibitemOpen
  \bibfield  {author} {\bibinfo {author} {\bibfnamefont {E.}~\bibnamefont
  {Babichev}}, \bibinfo {author} {\bibfnamefont {C.}~\bibnamefont {Deffayet}},\
  and\ \bibinfo {author} {\bibfnamefont {R.}~\bibnamefont {Ziour}},\ }\bibfield
   {title} {\bibinfo {title} {{The Vainshtein mechanism in the Decoupling Limit
  of massive gravity}},\ }\href {https://doi.org/10.1088/1126-6708/2009/05/098}
  {\bibfield  {journal} {\bibinfo  {journal} {JHEP}\ }\textbf {\bibinfo
  {volume} {0905}},\ \bibinfo {pages} {098}},\ \Eprint
  {https://arxiv.org/abs/0901.0393} {arXiv:0901.0393 [hep-th]} \BibitemShut
  {NoStop}%
\bibitem [{\citenamefont {Babichev}\ \emph {et~al.}(2010)\citenamefont
  {Babichev}, \citenamefont {Deffayet},\ and\ \citenamefont
  {Ziour}}]{Babichev:2010jd}%
  \BibitemOpen
  \bibfield  {author} {\bibinfo {author} {\bibfnamefont {E.}~\bibnamefont
  {Babichev}}, \bibinfo {author} {\bibfnamefont {C.}~\bibnamefont {Deffayet}},\
  and\ \bibinfo {author} {\bibfnamefont {R.}~\bibnamefont {Ziour}},\ }\bibfield
   {title} {\bibinfo {title} {{The Recovery of General Relativity in massive
  gravity via the Vainshtein mechanism}},\ }\href
  {https://doi.org/10.1103/PhysRevD.82.104008} {\bibfield  {journal} {\bibinfo
  {journal} {Phys.Rev.}\ }\textbf {\bibinfo {volume} {D82}},\ \bibinfo {pages}
  {104008} (\bibinfo {year} {2010})},\ \Eprint
  {https://arxiv.org/abs/1007.4506} {arXiv:1007.4506 [gr-qc]} \BibitemShut
  {NoStop}%
\bibitem [{\citenamefont {Chkareuli}\ and\ \citenamefont
  {Pirtskhalava}(2012)}]{Chkareuli:2011te}%
  \BibitemOpen
  \bibfield  {author} {\bibinfo {author} {\bibfnamefont {G.}~\bibnamefont
  {Chkareuli}}\ and\ \bibinfo {author} {\bibfnamefont {D.}~\bibnamefont
  {Pirtskhalava}},\ }\bibfield  {title} {\bibinfo {title} {{Vainshtein
  Mechanism In $\Lambda_3$ - Theories}},\ }\href
  {https://doi.org/10.1016/j.physletb.2012.05.030} {\bibfield  {journal}
  {\bibinfo  {journal} {Phys.Lett.}\ }\textbf {\bibinfo {volume} {B713}},\
  \bibinfo {pages} {99} (\bibinfo {year} {2012})},\ \Eprint
  {https://arxiv.org/abs/1105.1783} {arXiv:1105.1783 [hep-th]} \BibitemShut
  {NoStop}%
\bibitem [{\citenamefont {Babichev}\ and\ \citenamefont
  {Deffayet}(2013)}]{Babichev:2013usa}%
  \BibitemOpen
  \bibfield  {author} {\bibinfo {author} {\bibfnamefont {E.}~\bibnamefont
  {Babichev}}\ and\ \bibinfo {author} {\bibfnamefont {C.}~\bibnamefont
  {Deffayet}},\ }\bibfield  {title} {\bibinfo {title} {{An introduction to the
  Vainshtein mechanism}},\ }\href
  {https://doi.org/10.1088/0264-9381/30/18/184001} {\bibfield  {journal}
  {\bibinfo  {journal} {Class. Quant. Grav.}\ }\textbf {\bibinfo {volume}
  {30}},\ \bibinfo {pages} {184001} (\bibinfo {year} {2013})},\ \Eprint
  {https://arxiv.org/abs/1304.7240} {arXiv:1304.7240 [gr-qc]} \BibitemShut
  {NoStop}%
\bibitem [{\citenamefont {Deffayet}\ and\ \citenamefont
  {Jacobson}(2012)}]{Deffayet:2011rh}%
  \BibitemOpen
  \bibfield  {author} {\bibinfo {author} {\bibfnamefont {C.}~\bibnamefont
  {Deffayet}}\ and\ \bibinfo {author} {\bibfnamefont {T.}~\bibnamefont
  {Jacobson}},\ }\bibfield  {title} {\bibinfo {title} {{On horizon structure of
  bimetric spacetimes}},\ }\href
  {https://doi.org/10.1088/0264-9381/29/6/065009} {\bibfield  {journal}
  {\bibinfo  {journal} {Class.Quant.Grav.}\ }\textbf {\bibinfo {volume} {29}},\
  \bibinfo {pages} {065009} (\bibinfo {year} {2012})},\ \Eprint
  {https://arxiv.org/abs/1107.4978} {arXiv:1107.4978 [gr-qc]} \BibitemShut
  {NoStop}%
\bibitem [{\citenamefont {Berens}\ \emph {et~al.}(2022)\citenamefont {Berens},
  \citenamefont {Krauth},\ and\ \citenamefont {Rosen}}]{Berens:2021tzd}%
  \BibitemOpen
  \bibfield  {author} {\bibinfo {author} {\bibfnamefont {R.}~\bibnamefont
  {Berens}}, \bibinfo {author} {\bibfnamefont {L.}~\bibnamefont {Krauth}},\
  and\ \bibinfo {author} {\bibfnamefont {R.~A.}\ \bibnamefont {Rosen}},\
  }\bibfield  {title} {\bibinfo {title} {{Gravitational collapse in massive
  gravity in de Sitter spacetime}},\ }\href
  {https://doi.org/10.1103/PhysRevD.105.064057} {\bibfield  {journal} {\bibinfo
   {journal} {Phys. Rev. D}\ }\textbf {\bibinfo {volume} {105}},\ \bibinfo
  {pages} {064057} (\bibinfo {year} {2022})},\ \Eprint
  {https://arxiv.org/abs/2109.10411} {arXiv:2109.10411 [hep-th]} \BibitemShut
  {NoStop}%
\bibitem [{\citenamefont {Rosen}(2017)}]{Rosen:2017dvn}%
  \BibitemOpen
  \bibfield  {author} {\bibinfo {author} {\bibfnamefont {R.~A.}\ \bibnamefont
  {Rosen}},\ }\bibfield  {title} {\bibinfo {title} {{Non-Singular Black Holes
  in Massive Gravity: Time-Dependent Solutions}},\ }\href
  {https://doi.org/10.1007/JHEP10(2017)206} {\bibfield  {journal} {\bibinfo
  {journal} {JHEP}\ }\textbf {\bibinfo {volume} {10}},\ \bibinfo {pages}
  {206}},\ \Eprint {https://arxiv.org/abs/1702.06543} {arXiv:1702.06543
  [hep-th]} \BibitemShut {NoStop}%
\bibitem [{\citenamefont {de~Rham}\ \emph {et~al.}(2023)\citenamefont
  {de~Rham}, \citenamefont {Ko\ifmmode~\dot{z}\else \.{z}\fi{}uszek},
  \citenamefont {Tolley},\ and\ \citenamefont {Wiseman}}]{deRham:2023ngf}%
  \BibitemOpen
  \bibfield  {author} {\bibinfo {author} {\bibfnamefont {C.}~\bibnamefont
  {de~Rham}}, \bibinfo {author} {\bibfnamefont {J.}~\bibnamefont
  {Ko\ifmmode~\dot{z}\else \.{z}\fi{}uszek}}, \bibinfo {author} {\bibfnamefont
  {A.~J.}\ \bibnamefont {Tolley}},\ and\ \bibinfo {author} {\bibfnamefont
  {T.}~\bibnamefont {Wiseman}},\ }\bibfield  {title} {\bibinfo {title}
  {Dynamical formulation of ghost-free massive gravity},\ }\href
  {https://doi.org/10.1103/PhysRevD.108.084052} {\bibfield  {journal} {\bibinfo
   {journal} {Phys. Rev. D}\ }\textbf {\bibinfo {volume} {108}},\ \bibinfo
  {pages} {084052} (\bibinfo {year} {2023})}\BibitemShut {NoStop}%
\bibitem [{\citenamefont {de~Rham}\ \emph
  {et~al.}(2011{\natexlab{a}})\citenamefont {de~Rham}, \citenamefont
  {Gabadadze},\ and\ \citenamefont {Tolley}}]{deRham:2010kj}%
  \BibitemOpen
  \bibfield  {author} {\bibinfo {author} {\bibfnamefont {C.}~\bibnamefont
  {de~Rham}}, \bibinfo {author} {\bibfnamefont {G.}~\bibnamefont {Gabadadze}},\
  and\ \bibinfo {author} {\bibfnamefont {A.~J.}\ \bibnamefont {Tolley}},\
  }\bibfield  {title} {\bibinfo {title} {{Resummation of Massive Gravity}},\
  }\href {https://doi.org/10.1103/PhysRevLett.106.231101} {\bibfield  {journal}
  {\bibinfo  {journal} {Phys.Rev.Lett.}\ }\textbf {\bibinfo {volume} {106}},\
  \bibinfo {pages} {231101} (\bibinfo {year} {2011}{\natexlab{a}})},\ \Eprint
  {https://arxiv.org/abs/1011.1232} {arXiv:1011.1232 [hep-th]} \BibitemShut
  {NoStop}%
\bibitem [{\citenamefont {de~Rham}\ \emph
  {et~al.}(2011{\natexlab{b}})\citenamefont {de~Rham}, \citenamefont
  {Gabadadze},\ and\ \citenamefont {Tolley}}]{deRham:2011qq}%
  \BibitemOpen
  \bibfield  {author} {\bibinfo {author} {\bibfnamefont {C.}~\bibnamefont
  {de~Rham}}, \bibinfo {author} {\bibfnamefont {G.}~\bibnamefont {Gabadadze}},\
  and\ \bibinfo {author} {\bibfnamefont {A.~J.}\ \bibnamefont {Tolley}},\
  }\bibfield  {title} {\bibinfo {title} {{Helicity Decomposition of Ghost-free
  Massive Gravity}},\ }\href {https://doi.org/10.1007/JHEP11(2011)093}
  {\bibfield  {journal} {\bibinfo  {journal} {JHEP}\ }\textbf {\bibinfo
  {volume} {1111}},\ \bibinfo {pages} {093}},\ \Eprint
  {https://arxiv.org/abs/1108.4521} {arXiv:1108.4521 [hep-th]} \BibitemShut
  {NoStop}%
\bibitem [{\citenamefont {de~Rham}\ and\ \citenamefont
  {Gabadadze}(2010)}]{deRham:2010ik}%
  \BibitemOpen
  \bibfield  {author} {\bibinfo {author} {\bibfnamefont {C.}~\bibnamefont
  {de~Rham}}\ and\ \bibinfo {author} {\bibfnamefont {G.}~\bibnamefont
  {Gabadadze}},\ }\bibfield  {title} {\bibinfo {title} {{Generalization of the
  Fierz-Pauli Action}},\ }\href {https://doi.org/10.1103/PhysRevD.82.044020}
  {\bibfield  {journal} {\bibinfo  {journal} {Phys. Rev.}\ }\textbf {\bibinfo
  {volume} {D82}},\ \bibinfo {pages} {044020} (\bibinfo {year} {2010})},\
  \Eprint {https://arxiv.org/abs/1007.0443} {arXiv:1007.0443 [hep-th]}
  \BibitemShut {NoStop}%
\bibitem [{\citenamefont {Ondo}\ and\ \citenamefont
  {Tolley}(2013)}]{Ondo:2013wka}%
  \BibitemOpen
  \bibfield  {author} {\bibinfo {author} {\bibfnamefont {N.~A.}\ \bibnamefont
  {Ondo}}\ and\ \bibinfo {author} {\bibfnamefont {A.~J.}\ \bibnamefont
  {Tolley}},\ }\bibfield  {title} {\bibinfo {title} {{Complete Decoupling Limit
  of Ghost-free Massive Gravity}},\ }\href
  {https://doi.org/10.1007/JHEP11(2013)059} {\bibfield  {journal} {\bibinfo
  {journal} {JHEP}\ }\textbf {\bibinfo {volume} {11}},\ \bibinfo {pages}
  {059}},\ \Eprint {https://arxiv.org/abs/1307.4769} {arXiv:1307.4769 [hep-th]}
  \BibitemShut {NoStop}%
\bibitem [{\citenamefont {de~Rham}\ \emph {et~al.}(2017)\citenamefont
  {de~Rham}, \citenamefont {Deskins}, \citenamefont {Tolley},\ and\
  \citenamefont {Zhou}}]{deRham:2016nuf}%
  \BibitemOpen
  \bibfield  {author} {\bibinfo {author} {\bibfnamefont {C.}~\bibnamefont
  {de~Rham}}, \bibinfo {author} {\bibfnamefont {J.~T.}\ \bibnamefont
  {Deskins}}, \bibinfo {author} {\bibfnamefont {A.~J.}\ \bibnamefont
  {Tolley}},\ and\ \bibinfo {author} {\bibfnamefont {S.-Y.}\ \bibnamefont
  {Zhou}},\ }\bibfield  {title} {\bibinfo {title} {{Graviton Mass Bounds}},\
  }\href {https://doi.org/10.1103/RevModPhys.89.025004} {\bibfield  {journal}
  {\bibinfo  {journal} {Rev. Mod. Phys.}\ }\textbf {\bibinfo {volume} {89}},\
  \bibinfo {pages} {025004} (\bibinfo {year} {2017})},\ \Eprint
  {https://arxiv.org/abs/1606.08462} {arXiv:1606.08462 [astro-ph.CO]}
  \BibitemShut {NoStop}%
\bibitem [{\citenamefont {Ko{\.z}uszek}\ and\ \citenamefont
  {Wiseman}(2025)}]{Kozuszek:2024vyb}%
  \BibitemOpen
  \bibfield  {author} {\bibinfo {author} {\bibfnamefont {J.}~\bibnamefont
  {Ko{\.z}uszek}}\ and\ \bibinfo {author} {\bibfnamefont {T.}~\bibnamefont
  {Wiseman}},\ }\bibfield  {title} {\bibinfo {title} {{Well-posedness of
  minimal dRGT massive gravity}},\ }\href
  {https://doi.org/10.1007/JHEP06(2025)133} {\bibfield  {journal} {\bibinfo
  {journal} {JHEP}\ }\textbf {\bibinfo {volume} {06}},\ \bibinfo {pages}
  {133}},\ \Eprint {https://arxiv.org/abs/2410.19491} {arXiv:2410.19491
  [hep-th]} \BibitemShut {NoStop}%
\bibitem [{\citenamefont {Hinterbichler}\ and\ \citenamefont
  {Rosen}(2012)}]{Hinterbichler:2012cn}%
  \BibitemOpen
  \bibfield  {author} {\bibinfo {author} {\bibfnamefont {K.}~\bibnamefont
  {Hinterbichler}}\ and\ \bibinfo {author} {\bibfnamefont {R.~A.}\ \bibnamefont
  {Rosen}},\ }\bibfield  {title} {\bibinfo {title} {{Interacting Spin-2
  Fields}},\ }\href {https://doi.org/10.1007/JHEP07(2012)047} {\bibfield
  {journal} {\bibinfo  {journal} {JHEP}\ }\textbf {\bibinfo {volume} {1207}},\
  \bibinfo {pages} {047}},\ \Eprint {https://arxiv.org/abs/1203.5783}
  {arXiv:1203.5783 [hep-th]} \BibitemShut {NoStop}%
\bibitem [{\citenamefont {Deser}\ \emph {et~al.}(2014)\citenamefont {Deser},
  \citenamefont {Sandora}, \citenamefont {Waldron},\ and\ \citenamefont
  {Zahariade}}]{Deser:2014hga}%
  \BibitemOpen
  \bibfield  {author} {\bibinfo {author} {\bibfnamefont {S.}~\bibnamefont
  {Deser}}, \bibinfo {author} {\bibfnamefont {M.}~\bibnamefont {Sandora}},
  \bibinfo {author} {\bibfnamefont {A.}~\bibnamefont {Waldron}},\ and\ \bibinfo
  {author} {\bibfnamefont {G.}~\bibnamefont {Zahariade}},\ }\bibfield  {title}
  {\bibinfo {title} {{Covariant constraints for generic massive gravity and
  analysis of its characteristics}},\ }\href
  {https://doi.org/10.1103/PhysRevD.90.104043} {\bibfield  {journal} {\bibinfo
  {journal} {Phys. Rev. D}\ }\textbf {\bibinfo {volume} {90}},\ \bibinfo
  {pages} {104043} (\bibinfo {year} {2014})},\ \Eprint
  {https://arxiv.org/abs/1408.0561} {arXiv:1408.0561 [hep-th]} \BibitemShut
  {NoStop}%
\bibitem [{\citenamefont {Berezhiani}\ \emph
  {et~al.}(2013{\natexlab{b}})\citenamefont {Berezhiani}, \citenamefont
  {Chkareuli}, \citenamefont {de~Rham}, \citenamefont {Gabadadze},\ and\
  \citenamefont {Tolley}}]{Berezhiani:2013dca}%
  \BibitemOpen
  \bibfield  {author} {\bibinfo {author} {\bibfnamefont {L.}~\bibnamefont
  {Berezhiani}}, \bibinfo {author} {\bibfnamefont {G.}~\bibnamefont
  {Chkareuli}}, \bibinfo {author} {\bibfnamefont {C.}~\bibnamefont {de~Rham}},
  \bibinfo {author} {\bibfnamefont {G.}~\bibnamefont {Gabadadze}},\ and\
  \bibinfo {author} {\bibfnamefont {A.}~\bibnamefont {Tolley}},\ }\bibfield
  {title} {\bibinfo {title} {{Mixed Galileons and Spherically Symmetric
  Solutions}},\ }\href {https://doi.org/10.1088/0264-9381/30/18/184003}
  {\bibfield  {journal} {\bibinfo  {journal} {Class.Quant.Grav.}\ }\textbf
  {\bibinfo {volume} {30}},\ \bibinfo {pages} {184003} (\bibinfo {year}
  {2013}{\natexlab{b}})},\ \Eprint {https://arxiv.org/abs/1305.0271}
  {arXiv:1305.0271 [hep-th]} \BibitemShut {NoStop}%
\end{thebibliography}%

\section{Appendix: details of the harmonic formulation}

Here we give the explicit form of the tensors that enter the harmonic Einstein equation~\eqref{eq:harm};
we have,
\begin{widetext}
\be
A^{\sigma\alpha\rho\beta}_{\mu\nu} = \left[ \left( 2 \left( m^2 + \frac{M^2 [ e]}{2}  \right) - 1 \right) (e^{-1})^{\rho\beta} \delta_\nu^\alpha - M^2 \left( (e^{-1} )^{\rho\beta} e_\nu^{~\alpha}  + g^{\rho\beta} \delta_\nu^{~\alpha} \right) \right] \delta_\mu^\sigma + g^{\sigma\rho} \delta_\mu^\alpha (e^{-1})^\beta_\nu + ( \mu \leftrightarrow \nu )
\ee
and defining,
\be
\tilde{B}^{\sigma\rho\delta\alpha\beta\gamma}_{\mu\nu} &=& 
\left[ 
- \frac{1}{4} \delta^\sigma_\mu \delta^\alpha_\nu \eta^{\delta\gamma}
+ \frac{1}{8} (e^{-1})^\delta_{~\mu} (e^{-1})^\gamma_{~\nu} g^{\sigma\alpha}
+ \left( ( m^2 + \frac{M^2 [ e ]}{2}  - \frac{1}{2} ) (e^{-1})^{\alpha\gamma} - \frac{M^2}{2} g^{\alpha\gamma} \right) 
(e^{-1})^\delta_{~\mu} \delta^\sigma_\nu
\right] g^{\rho\beta} \nl 
&& \qquad \qquad - \frac{M^2}{2} (e^{-1})^\delta_{~\mu} \delta^\sigma_{~\nu} (e^{-1})^{\alpha\gamma} e^{\rho\beta} + ( \mu \leftrightarrow \nu )
\ee
\be
\tilde{C}^{\sigma\rho\delta\alpha\beta\gamma}_{\mu\nu} & = &
\frac{1}{2} ( e^{-1})^{\gamma\rho} g^{\sigma\alpha} (e^{-1})^\delta_{~\mu} \delta^\beta_{\nu} 
- \frac{1}{2} ( e^{-1})^{\beta\rho} g^{\gamma\alpha} (e^{-1})^\delta_{~\mu} \delta^\sigma_{\nu} 
+ \frac{1}{4} ( e^{-1} )^{\alpha\delta} (e^{-1})^{\rho\gamma} \delta^\sigma_\mu \delta^\beta_\nu \nl
&& + \left[ 
\left( \frac{1}{2} -  m^2 - \frac{M^2[e]}{2}  ) \right) (e^{-1})^{\sigma\gamma} (e^{-1})^{\rho\delta} + 
\frac{M^2}{2} \left( g^{\sigma\gamma} (e^{-1})^{\rho\delta} + (e^{-1})^{\sigma\gamma} g^{\rho\delta} \right)
\right] \delta^\alpha_\mu \delta^\beta_\nu \nl
&& + \left[ 
\left( \frac{1}{4} -  m^2 - \frac{M^2[e] }{2} ) \right) (e^{-1})^{\rho\gamma} (e^{-1})^{\beta\delta} + 
\frac{M^2}{2} \left( - g^{\beta\gamma} (e^{-1})^{\rho\delta}  + g^{\rho\beta} (e^{-1})^{\gamma\delta} + (e^{-1})^{\delta\beta} g^{\gamma\rho} \right)
\right] \delta^\sigma_\mu \delta^\alpha_\nu \nl
&& + \frac{M^2}{2} (e^{-1})^{\sigma\gamma} (e^{-1})^{\rho\delta} e^\beta_{~\mu} \delta^\alpha_{~\nu} 
+ \frac{M^2}{2} (e^{-1})^{\rho\beta} (e^{-1})^{\gamma\delta} e^\sigma_{~\mu} \delta^\alpha_{~\nu} + ( \mu \leftrightarrow \nu )
\ee
then,
\be
{B}^{\sigma\rho\delta\alpha\beta\gamma}_{\mu\nu} = \tilde{B}^{\sigma\rho\delta\alpha\beta\gamma}_{\mu\nu} + \tilde{C}^{\sigma\rho\delta[\alpha\beta]\gamma}_{\mu\nu} \; , \qquad
{C}^{\sigma\rho\delta\alpha\beta\gamma}_{\mu\nu} = \tilde{C}^{\sigma\rho\delta(\alpha\beta)\gamma}_{\mu\nu}
\ee
where we recall we have chosen units so that the graviton mass $\mu^2 = 1 = m^2 + M^2$.
For the scalar constraint~\eqref{eq:scalar} we first define,
\be
U^{\sigma\delta\alpha\gamma} &=&  - \frac{1}{8} g^{\sigma\alpha} \eta^{\delta\gamma} + \frac{1}{2} (e^{-1})^{\sigma\delta} (e^{-1})^{\alpha\gamma} - \frac{1}{4} (e^{-1})^{\sigma\gamma} (e^{-1})^{\alpha\delta}  \nl
 V^{\sigma\delta\alpha\gamma} &=&  \frac{1}{4} e^{\sigma\alpha} \eta^{\delta\gamma} + \frac{1}{8} (e^{-1})^{\gamma\delta} g^{\sigma\alpha} - \frac{1}{4} (e^{-1})^{\sigma\delta} g^{\alpha\gamma}- \frac{1}{4} (e^{-1})^{\alpha\gamma} g^{\sigma\delta} \nl
W^{\sigma\delta\alpha\gamma} &=& - \frac{1}{2} (e^{-1})^{\sigma\delta} (e^{-1})^{\alpha\gamma} + \frac{1}{4} (e^{-1})^{\sigma\gamma} (e^{-1})^{\alpha\delta} 
\ee
and then,
\be
&& S^{\sigma\rho\delta\alpha\beta\gamma} = \Big[
( m^2 + \frac{M^2 [e]}{2}  ) U^{\sigma\delta\alpha\gamma}  + M^2 V^{\sigma\delta\alpha\gamma} \Big] g^{\rho\beta} 
 + M^2 W^{\sigma\delta\alpha\gamma} e^{\rho\beta} \nonumber
\ee
\be
\tilde{M} = \frac{3}{2} m^4 \left( [e] - 4 \right)
+ \frac{m^2 M^2}{4} \left( 3 [e]^2 - 3 [e^2] - 6 [e]-12 \right) + \frac{M^4}{8} \left( [e]^3 - 3 [e][e^2] + 2 [e^3] - 6 [e] \right)
\ee
and for its time derivative,
\be
F^{\sigma\rho\delta\alpha\beta\gamma\epsilon\eta} &=&
- 2 (e^{-1})^{\epsilon(\rho} g^{\beta)\eta} \Big[ ( m^2 + \frac{M^2 [e]}{2}  ) U^{\sigma\delta\alpha\gamma}  + M^2 V^{\sigma\delta\alpha\gamma}
\Big]  - M^2 \left( 2 (e^{-1})^{\epsilon(\rho} e^{\beta)\eta} + g^{\epsilon(\rho} g^{\beta)\eta} \right) W^{\sigma\delta\alpha\gamma} \nl
&&
- ( m^2 + \frac{M^2 [e]}{2}  ) g^{\rho\beta} \left(  (e^{-1})^{\epsilon\sigma} U^{\eta\delta\alpha\gamma} + (e^{-1})^{\epsilon\alpha} U^{\eta\gamma\sigma\delta} \right) 
- \frac{1}{2} M^2 g^{\rho\beta} g^{\epsilon\eta} U^{\sigma\delta\alpha\gamma} \nl
&& \qquad + M^2 g^{\rho\beta} \left( 
- \frac{1}{2} \eta^{\delta\gamma} ( (e^{-1})^{\epsilon(\sigma} e^{\alpha)\eta} + \frac{1}{2} g^{\epsilon(\sigma} g^{\alpha)\eta} )
- \frac{1}{8} g^{\sigma\alpha} (e^{-1})^{\epsilon(\gamma} (e^{-1})^{\delta) \eta} - \frac{1}{4} (e^{-1})^{\gamma\delta} (e^{-1})^{\epsilon(\sigma} g^{\alpha)\eta}
\right) \nl
&& \qquad + M^2 g^{\rho\beta} \left( \frac{1}{2} g^{\alpha\gamma} (e^{-1})^{\delta\eta} (e^{-1})^{\sigma\epsilon} + (e^{-1})^{\sigma\delta} (e^{-1})^{\epsilon(\alpha} g^{\gamma)\eta} \right) \nl
&& \qquad - M^2 e^{\rho\beta} \left((e^{-1})^{\epsilon\sigma} W^{\eta\delta\alpha\gamma} + (e^{-1})^{\epsilon\alpha} W^{\eta\gamma\sigma\delta} \right) 
\ee
and finally,
\be
H^{\alpha\beta} = \frac{3 m^4}{2}  g^{\alpha\beta} - \frac{3 m^2 M^2}{2} \left( (1 - [e]) g^{\alpha\beta} + e^{\alpha\beta} \right)
- \frac{M^4}{8} \left( ( 6 + 3 [e^2] - 3 [e]^2 ) g^{\alpha\beta} + 6 [e] e^{\alpha\beta} - 6 e^\alpha_{~\rho} e^{\rho \beta} \right) \; .
\ee

\end{widetext}

\end{document}